\documentclass[a4paper]{article}
\usepackage{tipa}
\usepackage{siunitx}
\usepackage{comment}
\usepackage{INTERSPEECH2021}

\usepackage{microtype}[activate]
\newcommand{\eg}{e.\,g.\,}
\newcommand{\ie}{i.\,e.\,}

\newcommand{\cf}{{cf.\,}}

\usepackage{multirow}
\usepackage{cleveref}
\usepackage{xcolor}
\usepackage{soul}

\newcommand{\dicova}{\textsc{DiCOVA}}
\newcommand{\covid}{COVID-19}

\title{EIHW-MTG DiCOVA 2021 Challenge System Report}

\name{Adria Mallol-Ragolta$^{1}$, Helena Cuesta$^{2}$, Emilia G\'omez$^{2,3}$, and Bj\"orn W.\ Schuller$^{1,4}$}
\address{\fontsize{11}{11}\selectfont
  $^{1}$ EIHW -- Chair of Embedded Intelligence for Health Care \& Wellbeing, University of Augsburg, Germany\\
  $^{2}$ MTG -- Music Technology Group, Universitat Pompeu Fabra, Spain\\
  $^{3}$ Joint Research Centre, European Commission\\
  $^{4}$ GLAM -- Group on Language, Audio \& Music, Imperial College London, UK}
\email{adria.mallol-ragolta@informatik.uni-augsburg.de}

\begin{document}
\maketitle
\begin{abstract}
\noindent
This paper aims to automatically detect \covid\, patients by analysing the acoustic information embedded in coughs. \covid\, affects the respiratory system, and, consequently, respiratory-related signals have the potential to contain salient information for the task at hand. We focus on analysing the spectrogram representations of coughing samples with the aim to investigate whether \covid\, alters the frequency content of these signals. Furthermore, this work also assesses the impact of gender in the automatic detection of \covid. To extract deep learnt representations of the spectrograms, we compare the performance of a cough-specific, and a Resnet18 pre-trained Convolutional Neural Network (CNN). Additionally, our approach explores the use of contextual attention, so the model can learn to highlight the most relevant 
deep learnt features extracted by the CNN. We conduct our experiments on the dataset released for the Cough Sound Track of the \dicova\, 2021 Challenge. The best performance on the test set is obtained using the Resnet18 pre-trained CNN with contextual attention, which scored an Area Under the Curve (AUC) of 70.91 at 80\,\% sensitivity.

\end{abstract}


\section{System Description}

This paper presents our contribution to the Cough Sound Track of the \textit{Diagnosing COVID-19 using Acoustics} (\dicova) 2021 Challenge~\cite{Muguli21-DCD}. \Cref{sec:overview} introduces the suggested approaches. 
\Cref{sec:preprocessing} describes the pre-processing applied to the coughing samples, and details the representations extracted for their modelling, which is explained in \Cref{sec:classifier}. 
Finally, \Cref{sec:results} summarises the results obtained from the approaches presented.

\subsection{Methodology Overview}
\label{sec:overview}

The outbreak of the \textit{Coronavirus Disease 2019} (\covid) has dramatically stressed the health systems worldwide. Despite the vaccines, massive population screenings will still be needed to control the spread of this disease and its strains. Current medical diagnostic tools are time consuming, and burden public expenditures. Thus, there is an opportunity to develop new digital, diagnostic tools to improve the monitoring and the early detection of \covid\, at a large scale cost-effectively.

The symptomatology of \covid\, includes affectations in the respiratory system. Therefore, the acoustics of respiratory-related body signals, such as coughs, have a high potential to contain salient information that can be used towards the diagnosis of \covid\, patients. A standard approach would be the extraction of engineering-based handcrafted features. However, it is not clear yet which are the optimal features for this problem. 

In this work, we opt for analysing the spectrogram representations of the coughing signals with the aim to i) investigate whether \covid\, symptomatology alters the frequency content of the coughing signals, and ii) assess the impact of gender in the automatic detection of \covid\, patients. To extract salient information from the spectrograms, our approach relies on \textit{Convolutional Neural Networks} (CNNs) 
combined with \textit{Fully Connected} (FC) layers responsible for the classification of the embedded features learnt. Furthermore, our approach 
explores the use of 
contextual attention 
so the network learns to highlight the most relevant embedded features for the task at hand. 

\subsection{Data Preparation}
\label{sec:preprocessing}

This section 
details 
the data preparation stage of our approach, which has several steps: audio resampling, silence removal, feature extraction, data patch generation, and data augmentation.

\subsubsection{Audio resampling}
\label{sec:preprocessing:resampling}

The training and validation partitions of the \dicova\, Challenge dataset for the Cough Sound Track are composed of $1\,040$ audio recordings of different durations, ranging from ca. $1$\,sec up to $15$\,sec, with an average duration of $4.7$\,sec.
The provided audio files are originally sampled at $44.1$\,kHz. Our preliminary spectral analysis of a subset of the recordings revealed that a substantial amount of them does not have any frequency content above $8$\,kHz. We hypothesize a potential reason to explain this is the use of low-quality equipment by patients when recording their coughing samples, \eg, with mobile devices. Therefore, we resample all audio files to a common sampling rate of $8$\,kHz to account for the diversity of devices used for recording. Besides, the lack of frequency content above $8$\,kHz results in a dark patch in the spectrogram representations of the corresponding coughing samples, which creates noise in the training data.

\subsubsection{Sound Activity Detection}
\label{sec:preprocessing:sad}

Each audio sample in the dataset contains a sequence of coughs. 
A short amount of silence separates consecutive coughing samples within each 
sequence. We consider these silent regions to be irrelevant in the detection of \covid, and, therefore, we use a \textit{Sound Activity Detector} (SAD) to filter them out.
After the resampling step, the audio files are subsequently passed through a 
SAD based on the \textit{Root-Mean-Square} (RMS) value of the audio samples in the time domain.
We compute the RMS using the \texttt{librosa} 
Python library~\cite{McFee20_Librosa080}, and a frame length of $64$\,msec. 
We use min-max normalization to scale each audio file's RMS, and we discard all frames below a threshold of $0.1$ (set empirically). After the SAD step, we concatenate all frames above the threshold, and save the result as a new audio file for further processing.
As an additional experiment, we compared the RMS-based SAD to a SAD based on spectral flux, which detects abrupt changes in the spectral domain. Although a cough is an example of such a change, 
the preliminary exploration of the results using both methods showed that the RMS-based SAD worked better in this context.
%
Note that to assess the effectiveness of silence removal in the detection of \covid\, patients, our experiments use both the original, and the cough-only audio files. 
Details about the experiments performed and the results obtained are given in \Cref{sec:classifier} and \Cref{sec:results}, respectively.

\subsubsection{Feature Extraction and Patch Generation}
\label{sec:preprocessing:features}

Our approach uses the spectrogram as the input representation of the coughing samples. 
We use the \textit{Short-Time Fourier Transform} (STFT) function from \texttt{librosa} to calculate the spectrogram of each audio sample in the dataset, 
using a window size of $1\,024$ samples ($128$\,msec), and a hop length of $128$ samples ($16$\,msec).
With this configuration, we extract the spectrograms using different parameters to compare their impact on the final results: we compare spectrograms with a linear frequency scale to their corresponding logarithmic frequency scale, and two different colormaps, namely, \textit{viridis} and \textit{magma}.
Note that the colormap parameter is especially relevant because the spectrograms are exported as images of $256 \times 256$ pixels for further use.
The preliminary experiments conducted to assess the impact of these parameters, not reported in this work, were not conclusive enough. Nonetheless, analysing the trends in the results obtained, we decided to focus our investigation on the spectrogram representations of the coughing samples using the logarithmic frequency scale, and \textit{magma} as the colormap.

To overcome the different durations of the samples in the dataset, we fix the length of the coughing samples to be fed into the models. Specifically, we decide modelling the coughing samples using acoustic frames of $1$\,sec length. Hence, the last step of the data preparation stage is the segmentation of all spectrograms into $1$-second length patches with a $50$\,\% overlap. With this strategy, several patches from a single coughing sample are used for training the models. 

\begin{table}[t]
\caption{Summary comparing the performance of the models trained with a cough-specific CNN. The results reported on the validation partition are computed by averaging the performances obtained in each individual fold. The model that scored the highest AUC is highlighted.}
\label{table:scratchWOattention}
    \vspace{-1.5em}
    \begin{center}
    \resizebox{\columnwidth}{!}{
        \begin{tabular}{lccrrr}
            \toprule
            \multicolumn{1}{c}{\textbf{Models}} & \multicolumn{1}{c}{\textbf{Audio Files}} & \multicolumn{1}{c}{\textbf{Partition}} & \multicolumn{1}{c}{\textbf{AUC}} & \multicolumn{1}{c}{\textbf{Sensitivity}} & \multicolumn{1}{c}{\textbf{Specificity}} \\
            \midrule
            \multirow{4}{*}{Baseline} & \multirow{2}{*}{Original} & Val. & 63.14 & 81.60 & 34.09\\
            & & Test & 54.31 & 80.49 & -- \\
            \cmidrule{2-6}
            & \multirow{2}{*}{Cough-only} & Val. & 62.86 & 81.60 & 38.65 \\
            & & Test & 58.31 & 80.49 & --\\
            \midrule
            \multirow{4}{*}{\parbox{1.1cm}{Gender Based}} & \multirow{2}{*}{Original} & Val. & 64.88 & 80.80 & 39.90\\
            & & Test & 52.88 & 80.49 & --\\
            \cmidrule{2-6}
            & \multirow{2}{*}{Cough-only} & Val. & 65.32 & 82.40 & 38.86\\
            & & Test & \textbf{59.04} & 80.49 & --\\
            \midrule
            \multirow{4}{*}{\parbox{1.1cm}{Gender Specific}} & \multirow{2}{*}{Original} & Val. & 58.91 & 83.20 & 33.89\\
            & & Test & 52.88 & 80.49 & -- \\
            \cmidrule{2-6}
            & \multirow{2}{*}{Cough-only} & Val. & 62.86 & 80.80 & 42.59 \\
            & & Test & 49.92 & 80.49 & --\\
            \bottomrule
        \end{tabular}
    }
    \end{center}
    \vspace{-2em}
\end{table}

\subsubsection{Data Augmentation}
\label{sec:preprocessing:augmentation}

The training dataset is unbalanced in terms of positive and negative examples: from the $1\,040$ samples in both the train and the validation partitions, $965$ are negative (healthy patients), 
while only $75$ are positive (\covid\, patients). 
This difference also impacts the number of spectrograms generated from the patches of the coughing samples defined. 
Thus, we increase the number of positive examples via replication, \ie, 
including copies of the positive spectrograms 
to balance the dataset.
We considered other forms of augmentation, such as filtering or additive noise.
However, since it is not clear yet which kind of information from the audio is relevant for the task at hand, we decided not to alter the 
acoustic content in any way.
Although replication introduces redundancy in the training set, we believe it is useful when the number of positive and negative examples significantly differs. 

\subsection{Classifier Description}
\label{sec:classifier}

This section presents the neural networks used to model the coughing samples to detect
\covid\, patients. While \Cref{sec:netArchitecture} describes the architecture of the networks implemented, \Cref{sec:genderAwareness} details the procedure for our networks to consider gender information. 

\begin{table}[t]
\caption{Summary comparing the performance of the models trained with a cough-specific CNN and contextual attention. The results reported on the validation partition are computed by averaging the performances obtained in each individual fold. The model that scored the highest AUC is highlighted.}
\label{table:scratchWattention}
    \vspace{-1.5em}
    \begin{center}
    \resizebox{\columnwidth}{!}{
        \begin{tabular}{lccrrr}
            \toprule
            \multicolumn{1}{c}{\textbf{Models}} & \multicolumn{1}{c}{\textbf{Audio Files}} & \multicolumn{1}{c}{\textbf{Partition}} & \multicolumn{1}{c}{\textbf{AUC}} & \multicolumn{1}{c}{\textbf{Sensitivity}} & \multicolumn{1}{c}{\textbf{Specificity}} \\
            \midrule
            \multirow{4}{*}{Baseline} & \multirow{2}{*}{Original} & Val. & 60.08 & 83.20 & 31.19 \\
            & & Test & 56.22 & 80.49 & -- \\
            \cmidrule{2-6}
            & \multirow{2}{*}{Cough-only} & Val. & 63.00 & 83.20 & 34.61 \\
            & & Test & 58.34 & 80.49 & -- \\
            \midrule
            \multirow{4}{*}{\parbox{1.1cm}{Gender Based}} & \multirow{2}{*}{Original} & Val. & 65.52 & 81.60 & 45.49 \\
            & & Test & 53.69 & 80.49 & -- \\
            \cmidrule{2-6}
            & \multirow{2}{*}{Cough-only} & Val. & 63.39 & 84.00 & 39.17\\
            & & Test & \textbf{61.62} & 80.49 & -- \\
            \midrule
            \multirow{4}{*}{\parbox{1.1cm}{Gender Specific}} & \multirow{2}{*}{Original} & Val. & 58.08 & 81.60 & 34.40\\
            & & Test & 50.77 & 80.49 & --\\
            \cmidrule{2-6}
            & \multirow{2}{*}{Cough-only} & Val. & 62.21 & 84.00 & 39.59 \\
            & & Test & 60.87 & 80.49 & -- \\
            \bottomrule
        \end{tabular}
    }
    \end{center}
    \vspace{-2em}
\end{table}

\subsubsection{Network Architectures}
\label{sec:netArchitecture}

The networks trained to detect \covid\, from coughing samples are composed of two main blocks: the first one extracts embedded representations from the input spectrograms, while the second one focuses on the classification of the embedded features depending on whether they belong to healthy or \covid\, patients. For the latter, we employ two FC layers with $128$ and $2$ output neurons, respectively. While the first layer uses \textit{Rectified Linear Unit} (ReLU) as the activation function, the second one uses Softmax, so that the network outputs can be interpreted as probability scores.

As our networks' inputs are spectrograms, the extraction of the embedded representations is implemented using CNNs. Specifically, we compare the performance of a cough-specific CNN trained from scratch with the performance of a Resnet18 pre-trained CNN~\cite{He16-DRL}. The cough-specific CNN is implemented with two convolutional blocks with 32 and 64 channels, respectively, a square kernel of $3 \times 3$, and a stride of $1$. Both blocks implement batch normalisation, and use ReLU as the activation function. While the first block includes a $2 \times 2$ max-pooling, the second one uses adaptive average pooling, so the learnt feature map has a dimension of $2 \times 2$.

To highlight the salient information from the embedded representations learnt, we include a contextual attention mechanism (adapted from~\cite{Yang16-HAN} and~\cite{Mallol19-AHA}) between the two blocks of the network. Representing the deep features learnt as $\boldsymbol{h}$, the contextual attention mechanism is mathematically defined as follows:
\begin{equation}
    \boldsymbol{u} = \tanh(\mathbf{W} \boldsymbol{h} + \mathbf{b}),
\end{equation}
\begin{equation}
    \boldsymbol{\alpha} = \frac{\exp\left(\boldsymbol{u}^T \mathbf{u_c}\right)}{\sum \exp\left(\boldsymbol{u}^T \mathbf{u_c}\right)},
\end{equation}
\begin{equation}
    \boldsymbol{\tilde{h}} = \boldsymbol{\alpha} \boldsymbol{h}, 
\end{equation}
where $\mathbf{W}$, $\mathbf{b}$, and $\mathbf{u_c}$ are parameters to be learnt by the network. The parameter $\mathbf{u_c}$ can be interpreted as the context vector. The attention-based representation obtained $\boldsymbol{\tilde{h}}$ is then fed into the FC layers for classification. 

\subsubsection{Gender Awareness}
\label{sec:genderAwareness}

Assessing the impact of gender in the automatic detection of \covid\, patients is also one of this work's goals.
To address this aspect, we explore three different network configurations.
The first one does not consider any gender information, and is used as a baseline for our experiments. The second one, referred to as gender-based models in our experiments, includes an encoded representation of the patients' gender, which is concatenated with the deep learnt features extracted. Both are fed into the FC layers of the network. The third and last configuration, referred to as gender-specific models in our experiments, trains gender-specific models, so female and male coughs are analysed with models trained using samples from patients of the same gender. 

\begin{table}[t]
\caption{Summary comparing the performance of the models trained with the pre-trained Resnet18 CNN. The results reported on the validation partition are computed by averaging the performances obtained in each individual fold. The model that scored the highest AUC is highlighted.}
\label{table:resnetWOattention}
    \vspace{-1.5em}
    \begin{center}
    \resizebox{\columnwidth}{!}{
        \begin{tabular}{lccrrr}
            \toprule
            \multicolumn{1}{c}{\textbf{Models}} & \multicolumn{1}{c}{\textbf{Audio Files}} & \multicolumn{1}{c}{\textbf{Partition}} & \multicolumn{1}{c}{\textbf{AUC}} & \multicolumn{1}{c}{\textbf{Sensitivity}} & \multicolumn{1}{c}{\textbf{Specificity}} \\
            \midrule
            \multirow{4}{*}{Baseline} & \multirow{2}{*}{Original} & Val. & 62.32 & 81.60 & 33.16 \\
            & & Test & 68.43 & 80.49 & --\\
            \cmidrule{2-6}
            & \multirow{2}{*}{Cough-only} & Val. & 53.55 & 81.60 & 16.48 \\
            & & Test & 54.55 & 80.49 & -- \\
            \midrule
            \multirow{4}{*}{\parbox{1.1cm}{Gender Based}} & \multirow{2}{*}{Original} & Val. & 64.35 & 82.40 & 38.24 \\
            & & Test & \textbf{68.95} & 80.49 & -- \\
            \cmidrule{2-6}
            & \multirow{2}{*}{Cough-only} & Val. & 55.33 & 84.00 & 18.45 \\
            & & Test & 52.31 & 82.93 & -- \\
            \midrule
            \multirow{4}{*}{\parbox{1.1cm}{Gender Specific}} & \multirow{2}{*}{Original} & Val. & 60.00 & 84.00 & 27.05\\
            & & Test & 51.94 & 80.49 & --\\
            \cmidrule{2-6}
            & \multirow{2}{*}{Cough-only} & Val. & 57.58 & 81.60 & 26.53\\
            & & Test & 58.62 & 97.56 & -- \\
            \bottomrule
        \end{tabular}
    }
    \end{center}
    \vspace{-2em}
\end{table}

\subsection{Results}
\label{sec:results}

All models are trained to minimise the Categorical Cross-Entropy Loss, using Adam as the optimiser with a fixed learning rate of $1e^{-3}$. Network parameters are updated in batches of $32$ samples, and trained during a maximum of $100$ epochs. We implement an early stopping mechanism to stop training when the validation loss does not improve for ten consecutive epochs. To assess the models, we follow a $5$-fold cross-validation approach, as defined by the Challenge organisers. Each fold is trained during a specific number of epochs. Therefore, when modelling all training material and to prevent overfitting, the training epochs are determined by computing the median of the training epochs processed in each fold.

As described in \Cref{sec:preprocessing:features}, several fix-length spectrograms can be extracted from a single coughing sample. Thus, at inference time, several probability scores can be predicted for a single sample. To overcome this issue, we compute the probability of a specific sample to belong to a \covid\, patient as the median of the probabilities inferred from the corresponding spectrograms. The results obtained when assessing the models trained using a cough-specific CNN without and with contextual attention are summarised in \Cref{table:scratchWOattention,table:scratchWattention}, respectively. The results obtained when assessing the models trained using the pre-trained Resnet18 CNN without and with contextual attention are compiled in \Cref{table:resnetWOattention,table:resnetWattention}, respectively.

\begin{table}[t]
\caption{Summary comparing the performance of the models trained with the pre-trained Resnet18 CNN and contextual attention. The results reported on the validation partition are computed by averaging the performances obtained in each individual fold. The model that scored the highest AUC is highlighted.}
\label{table:resnetWattention}
    \vspace{-1.5em}
    \begin{center}
    \resizebox{\columnwidth}{!}{
        \begin{tabular}{lccrrr}
            \toprule
            \multicolumn{1}{c}{\textbf{Models}} & \multicolumn{1}{c}{\textbf{Audio Files}} & \multicolumn{1}{c}{\textbf{Partition}} & \multicolumn{1}{c}{\textbf{AUC}} & \multicolumn{1}{c}{\textbf{Sensitivity}} & \multicolumn{1}{c}{\textbf{Specificity}} \\
            \midrule
            \multirow{4}{*}{Baseline} & \multirow{2}{*}{Original} & Val. & 62.97 & 81.60 & 37.41 \\
            & & Test & 69.17 & 80.49 & --\\
            \cmidrule{2-6}
            & \multirow{2}{*}{Cough-only} & Val. & 54.93 & 82.40 & 21.55\\
            & & Test & 52.56 & 80.49 & -- \\
            \midrule
            \multirow{4}{*}{\parbox{1.1cm}{Gender Based}} & \multirow{2}{*}{Original} & Val. & 61.59 & 80.80 & 31.19\\
            & & Test & 69.89 & 80.49 & -- \\
            \cmidrule{2-6}
            & \multirow{2}{*}{Cough-only} & Val. & 55.38 & 80.80 & 24.56\\
            & & Test & 54.34 & 80.49 & --\\
            \midrule
            \multirow{4}{*}{\parbox{1.1cm}{Gender Specific}} & \multirow{2}{*}{Original} & Val. & 62.01 & 83.20 & 36.27\\
            & & Test & \textbf{70.91} & 80.49 & -- \\
            \cmidrule{2-6}
            & \multirow{2}{*}{Cough-only} & Val. & 59.56 & 81.60 & 35.23 \\
            & & Test & 59.01 & 80.49 & -- \\
            \bottomrule
        \end{tabular}
    }
    \end{center}
    \vspace{-2em}
\end{table}
One of our experiments' main insights is that models that incorporate gender information outperform the baseline model in most cases.
In this task, the gender of the patient is especially relevant: the vocal apparatus has a different shape and size for males and females, which results in significant differences both in the timbre and frequency range of the respiratory-related signals. 
We obtained the best performance with the gender-based model in three of the four scenarios we present in this paper. 
Besides, it is worth mentioning that although in \Cref{table:resnetWattention} the best performance is achieved with the gender-specific model, the corresponding \textit{Area Under the Curve} (AUC) for the gender-based model is only $1$\,\% lower, suggesting that they have an equivalent performance.
Since both models incorporate gender information, these results lead us to conclude that the gender of the patient is a very relevant feature for this task.

When we compare the results between cough-only and original audio files, we observe a clear difference: interestingly, 
the best performances using cough-only audio files on the test set were obtained with the cough-specific CNN (\cf \Cref{table:scratchWOattention,table:scratchWattention}), 
while original audio files scored the highest AUC using the pre-trained ResNet18 CNN (\cf \Cref{table:resnetWOattention,table:resnetWattention}).
One potential reason behind these differences is that ResNet18 is a pre-trained network for image classification and not directly related to acoustics. 
In general, images can be quite heterogeneous, \ie, they commonly have several elements, separated by edges, with higher gradients to be captured by the CNN. In our spectrogram images, part of the edges appear at the beginning and end of the silent regions 
because of the abrupt change in the frequency content. 
The cough-only audio files do not contain these silent regions, and, therefore, they become much more homogeneous. 
In contrast, the original audio files have more edges, and they would be more suitable for such a pre-trained network.

\section{Acknowledgement}

\noindent
This project has received funding from the European Union's Horizon 2020 research and innovation programme under grant agreements No.\,826506 (sustAGE) and No.\,770376 (TROMPA).
Further funding has been received from the FI Predoctoral Grant 2018FI-B01015 from AGAUR, Generalitat de Catalunya.

\bibliographystyle{IEEEtran}

\bibliography{mybib}

\end{document}